\documentclass{emulateapj}

\usepackage{amsmath}
\usepackage{epsfig}
\usepackage{amssymb}
\usepackage{graphicx}
\usepackage{color}
\usepackage{natbib}

\def\gtaprx {\lower .1ex\hbox{\rlap{\raise .6ex\hbox{\hskip .3ex
	{\ifmmode{\scriptscriptstyle >}\else
		{$\scriptscriptstyle >$}\fi}}}
	\kern -.4ex{\ifmmode{\scriptscriptstyle \sim}\else
		{$\scriptscriptstyle\sim$}\fi}}}
\def\ltaprx {\lower .1ex\hbox{\rlap{\raise .6ex\hbox{\hskip .3ex
	{\ifmmode{\scriptscriptstyle <}\else
		{$\scriptscriptstyle <$}\fi}}}
	\kern -.4ex{\ifmmode{\scriptscriptstyle \sim}\else
		{$\scriptscriptstyle\sim$}\fi}}}

\newcommand{\cutt}[1]{\textcolor{blue}{}}

\newcommand{\Ms}{{\ensuremath{M_{\odot} }}}

\begin{document}

\title{Finding Direct-Collapse Black Holes at Birth}

\author{Daniel J. Whalen,\altaffilmark{1,2} Marco Surace,\altaffilmark{1} Carla Bernhardt,\altaffilmark{3} Erik Zackrisson,\altaffilmark{4} Fabio Pacucci, \altaffilmark{5,6} Bodo L. Ziegler,\altaffilmark{7} and Michaela Hirschmann\altaffilmark{8}}

\altaffiltext{1}{Institute of Cosmology and Gravitation, University of Portsmouth, Portsmouth PO1 3FX, UK}
\altaffiltext{2}{Ida Pfeiffer Professor, University of Vienna, Department of Astrophysics, Tuerkenschanzstrasse 17, 1180, Vienna, Austria}
\altaffiltext{3}{Universit\"at Heidelberg, Institut f\"ur Theoretische Astrophysik, Albert-Ueberle-Str. 2, 69120 Heidelberg, Germany}
\altaffiltext{4}{Observational Astrophysics, Department of Physics and Astronomy, Uppsala University, Box 516, SE-751 20 Uppsala, Sweden}
\altaffiltext{5}{Black Hole Initiative, Harvard University, Cambridge, MA 02138, USA} 
\altaffiltext{6}{Center for Astrophysics $\vert$ Harvard \& Smithsonian, Cambridge, MA 02138, USA}
\altaffiltext{7}{University of Vienna, Department of Astrophysics, Tuerkenschanzstrasse 17, 1180, Vienna, Austria}
\altaffiltext{8}{DARK, Niels Bohr Institute, University of Copenhagen, Lyngbyvej 2, DK-2100 Copenhagen, Denmark}

\begin{abstract}

Direct-collapse black holes (DCBHs) are currently one of the leading contenders for the origins of the first quasars in the universe, over 300 of which have now been found at $z >$ 6.  But the birth of a DCBH in an atomically-cooling halo does not by itself guarantee it will become a quasar by $z \sim$ 7, the halo must also be located in cold accretion flows or later merge with a series of other gas-rich halos capable of fueling the BH's rapid growth.  Here, we present near infrared luminosities for DCBHs born in cold accretion flows in which they are destined to grow to 10$^9$ \Ms\ by $z \sim$ 7.  Our observables, which are derived from cosmological simulations with radiation hydrodynamics with Enzo, reveal that DCBHs could be found by the {\em James Webb Space Telescope} at $z \lesssim$ 20 and strongly-lensed DCBHs might be found in future wide-field surveys by {\em Euclid} and the Wide-Field Infrared Space Telescope at $z \lesssim$ 15.

\end{abstract}


\maketitle

\section{Introduction}

DCBHs may be the origins of the first quasars in the universe \citep[e.g.,][]{mort11,ban18,mats19}.  They are thought to form in primordial halos immersed in either strong Lyman-Werner (LW) UV fluxes or highly supersonic baryon streaming flows, either of which can prevent them from forming primordial (or Pop III) stars until they reach masses of 10$^7$ - 10$^8$ \Ms\ and virial temperatures of $\sim$ 10$^4$ K that trigger rapid atomic H cooling \citep{sb19,ivh20}.  Atomic cooling causes gas to collapse at rates of up to $\sim$ 1 \Ms\ yr$^{-1}$, forming an accretion disk that builds up a single, supermassive star at its center (e.g., \citealt{rh09b,latif13a} -- although binaries or even small multiples are now thought to be possible; \citealt{latif20a}). 

Stellar evolution models show that these stars can reach masses of a few 10$^5$ \Ms\ before  collapsing to DCBHs via the general relativistic instability \citep{um16,tyr17,hle18,hle17}, although a few for which accretion has shut down have been found to explode as highly energetic thermonuclear supernovae \citep{wet13b,wet13a,jet13a,chen14b}. DCBHs are currently the leading candidates for the seeds of the first supermassive black holes (SMBHs) because they are born with large masses in high densities in halos that can retain their fuel supply, even when heated by X-rays (\citealt{jet13} -- see \citealt{rosa17,titans} for recent reviews).  In contrast, while Pop III star BHs in principle can reach 10$^9$ \Ms\ with periodic episodes of super- or hyper-Eddington accretion \citep{vsd15,pez16}, their environments are hostile to such growth \citep{wan04,wf12,srd18}. 

The formation of a DCBH does not guarantee it will become a 10$^9$ \Ms\ quasar by $z >$ 6 because a large gas reservoir is also needed for its sustained growth.  Its host halo must either lie at the nexus of cold accretion flows \citep[e.g.,][]{dm12,yue14} or undergo a series of mergers with other gas-rich halos capable of fueling its rapid growth \citep[e.g.,][]{li07}.  Atomically-cooled halos fed by cold streams are more turbulent than other halos because they can reach masses greater than 10$^{12}$ \Ms\ by $z \sim 7$ \citep{smidt18}.  DCBHs born in such environments can thus grow more rapidly than in other halos.

What are the prospects for detecting DCBHs, and thus the birth of the first quasars?  Using one-dimensional (1D) radiation hydrodynamical models, \citet{pac15} predicted that DCBHs could be detected by the {\em James Webb Space Telescope} ({\em JWST}) in the near infrared (NIR) at $z \sim$ 25 and by the Advanced Telescope for High-Energy Astrophysics (ATHENA) at $z \sim$ 15.  \citet{nat17} used such models to develop criteria for distinguishing the host galaxies of DCBHs from those of other SMBH seeds at $z \sim$ 10, showing that {\em JWST} can distinguish between seeding mechanisms at this redshift.  \citet{bar18} post processed cosmological simulations of DCBH birth in atomically-cooling halos with Monte Carlo radiative transfer to produce synthetic BH spectra.  Their models included X-rays from the BH and star formation and supernova feedback in its host halo.  They found similar limits for DCBH detections in the NIR as \citet{pac15} and \citet{nat17} with some spectral differences owing in part to star formation in their halo.

But these models either assume idealized halo profiles or halos that are not fed by cold streams or later grow to large masses.  Like the supermassive stars from which they form, DCBHs are deeply imbedded in these flows, which can heavily reprocess radiation from the BH in ways that have not been considered in previous studies, changing their rest frame spectra and NIR luminosities today \citep[see also][for how radiative transfer effects deep in these halos can affect the dynamics of their flows]{ge17,aaron17,wg20}.  Here, we calculate NIR AB magnitudes for a DCBH at birth in the flows in which it grows into a quasar by $z \sim$ 7.  Rather than assuming a grid of accretion rates for the BH, ours are an emergent feature of a cosmological simulation.  Our models capture the anisotropy of X-ray breakout into the early intergalactic medium (IGM) and how it affects their detection today.  In Section~2 we describe our calculations and examine DCBH spectra and AB magnitudes for a variety of {\em JWST}, {\em Euclid} and Wide-Field Infrared Survey Telescope (WFIRST) bands in Section~3.  We conclude in Section~4.

\section{Numerical Method}

\begin{figure*} 
\begin{center}
\begin{tabular}{ccc}
\includegraphics[width=0.40\textwidth]{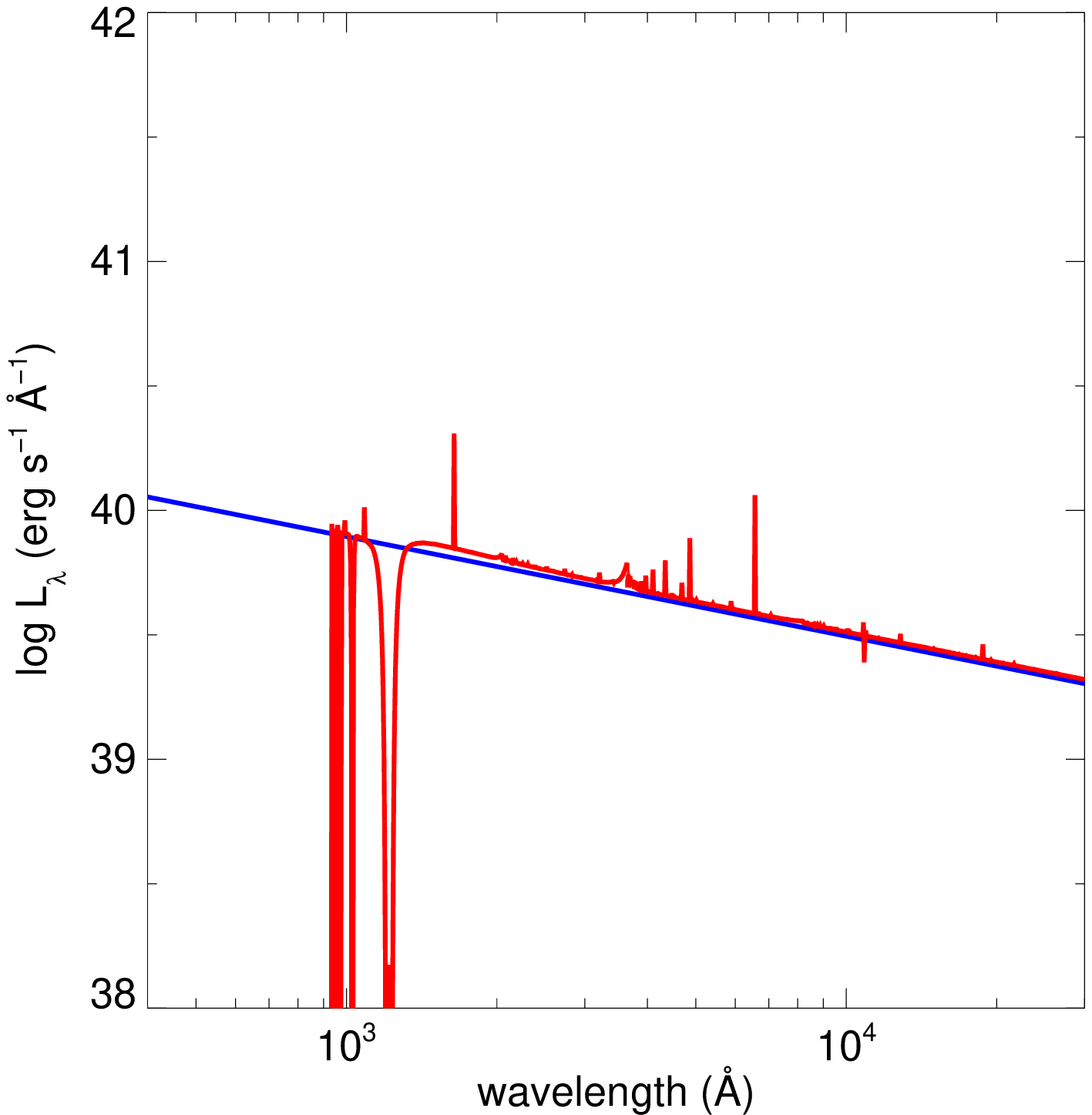} 
\includegraphics[width=0.55\textwidth]{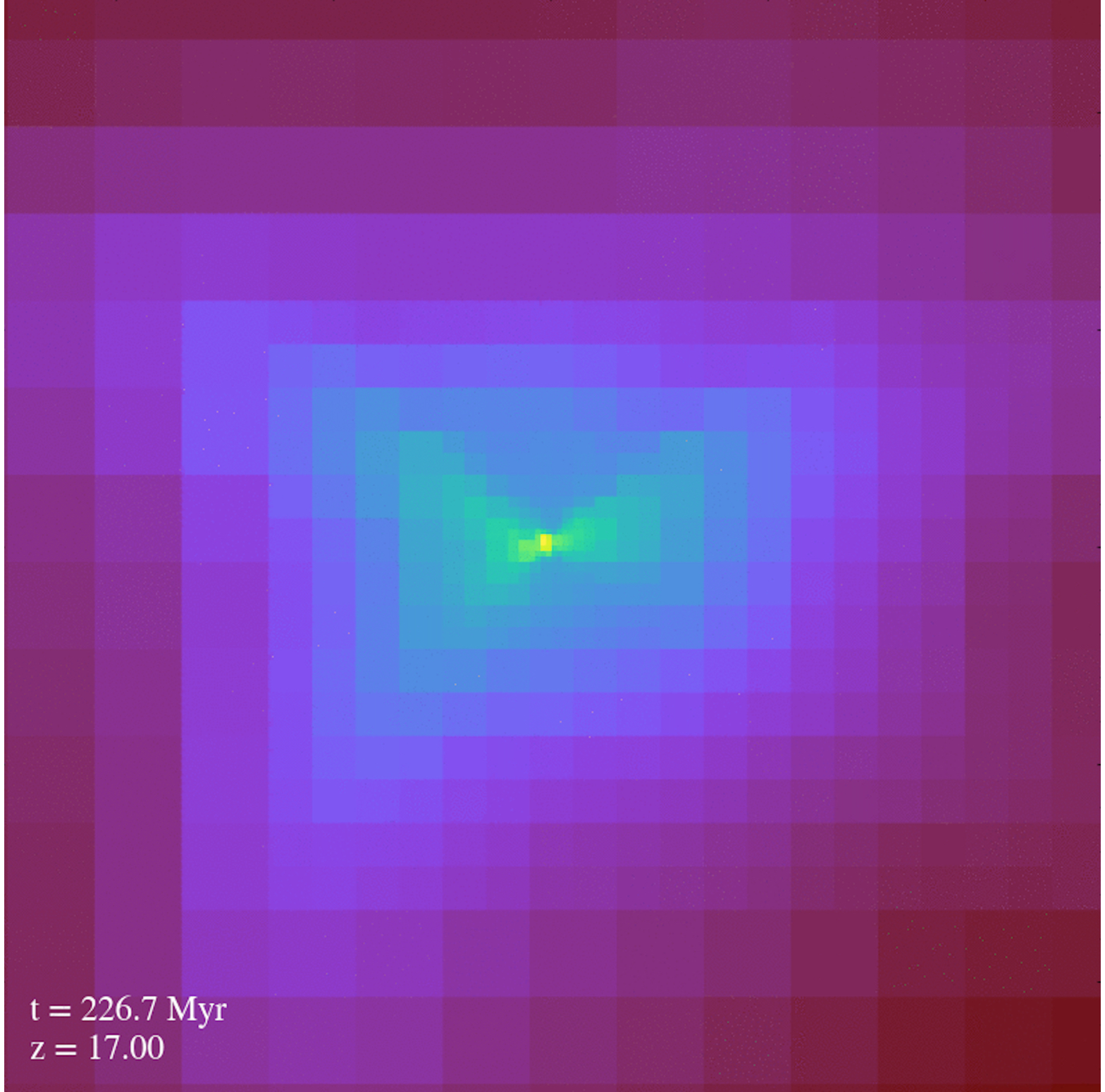}  
\end{tabular}
\end{center}
\caption{Birth of a DCBH at $z = $ 19.  Left panel: rest frame spectra for the DCBH at $z =$ 19 before (blue) and after (red) reprocessing by its host halo.  Right panel: ionized H fractions in the vicinity of the BH at $z =$ 17.  The image is 30 kpc proper on a side.}
\vspace{0.1in}
\label{fig:DCBH} 
\end{figure*}

We first extract luminosities and H II region profiles for the DCBH from \citet{smidt18}, which was done with the Enzo adaptive mesh refinement (AMR) cosmology code \citep{enzo}. They are then post processed with Cloudy \citep{cloudy17} to obtain rest frame BH spectra.  These spectra are then cosmologically redshifted and dimmed, corrected by absorption by the neutral IGM at $z >$ 6, and convolved with telescope filter functions to compute AB magnitudes in a variety of NIR bands as a function of source redshift.

\subsection{Enzo Model}

In the \citet{smidt18} Enzo simulation, a halo in a 100 $h^{-1}$ Mpc box reaches a mass of $3\times10^8$ \Ms\ at $z =$ 19.2 and begins to atomically cool.  At this redshift X-rays from a 10$^5$ \Ms\ DCBH are turned on in the halo, which later grows to $1.2\times 10^{12}$ \Ms\ by $z =$ 7.1 by accretion rather than major mergers.  The X-rays were propagated with MORAY \citep{moray}, which is self-consistently coupled to hydrodynamics and nine-species nonequilibrium primordial gas chemistry in Enzo.  The simulation included radiation pressure on gas due to photoionizations, Compton heating by X-rays, and primordial gas cooling. 
 
A maximum of 10 levels of adaptive mesh refinement produced a resolution of 35 pc (comoving), which was sufficient to resolve the gas flows and radiation transport deep in the halo.  The DCBH was represented by a sink particle with 1 keV X-rays and an alpha disk model for accretion rates to approximate the transport of angular momentum out of the disk on subgrid scales.  Although Pop II and III star formation, stellar winds, and ionizing UV and supernovae due to stars were turned on at the same time as X-rays from the BH, no stars formed in the short times we examine the DCBH here so its host halo is still free of metals.  An image of the H II region of the DCBH at $z =$ 17 is shown in the right panel of Figure~\ref{fig:DCBH}. 

\subsection{Cloudy Spectra}

To compute DCBH spectra we port spherically-averaged density and temperature profiles of the H II region of the BH from Enzo to Cloudy.  They are tabulated in 33 bins that are uniformly partitioned in log radius and extend to the outer layers of the H II region where the temperature of the gas has fallen below 10$^4$ K ($\sim$ 30 kpc).  Each radial bin, or shell, constitutes a single Cloudy model in which densities and temperatures are assumed to be constant.  The spectrum emerging from the outer surface of one shell is calculated and then used as the incident spectrum of the next shell.  The spectrum emerging from the outermost shell of the H II region is taken to be the rest frame spectrum of the quasar.  Although our 1D profiles smoothen out anisotropies in angle due to cosmological structures, they capture their average effects on AB magnitudes.  The actual magnitudes could be brighter along some lines of sight than others because of these anisotropies.  

We apply the default Cloudy broken power-law spectrum to the lower face of the innermost shell because of its similarity to that used in \citet{pac15}, where $F_{\nu} \propto \nu^{\alpha}$ and $\alpha = -2$ for h$\nu >$ 50 keV (2.48 $\times$ 10$^{-5}$ $\mu$m), $\alpha = -1.6$ for 50 keV $>$ h$\nu$ $>$ 0.124 eV (10 $\mu$m), and $\alpha = 5/2$ above 10 $\mu$m.  It is normalized to the bolometric luminosity of the DCBH. Coronal equilibrium is assumed, in which the gas is collisionally ionized.  We require Cloudy to use the temperatures Enzo calculates for the H II region instead of inferring them from the spectrum and luminosity of the BH and its surrounding density field because they take into account cooling due to nonequilibrium primordial gas chemistry in cosmological flows.  How we compute AB magnitudes from rest frame Cloudy spectra is described in detail in \citet{sur18a}.

\section{Detecting DCBHs}

We show rest frame spectra for the DCBH at $z =$ 19 before and after reprocessing by the  halo in the left panel of Figure~\ref{fig:DCBH}.  It has a bolometric luminosity of 2.42 $\times$ 10$^{44}$ erg s$^{-1}$ corresponding to an accretion rate of 0.85 $L_{\mathrm{Edd}}$. There is a conspicuous lack of metal lines  in the emergent spectrum because X-rays from the BH have not yet triggered star formation. Strong Ly$\alpha$ absorption is evident at 1216 \AA\ as is continuum absorption below 912 \AA\ due to the ionization of H.  Additional absorption features due to ionization of He I and He II are visible at 504 \AA\ and 227 \AA, respectively. Several prominent He emission lines are superimposed on the continuum absorption below 912 \AA.  There are H$\alpha$ and weak Paschen series lines at 6560 \AA\ and  12800 \AA. Unlike the spectrum of the cool, red progenitor star \citep{sur18a}, there is a lack of continuum absorption above and below 16500 \AA\ due to H$^-$ bound-bound and bound-free opacity in the DCBH spectrum because it is destroyed by radiation from the BH.  

\subsection{NIR Magnitudes}

\begin{figure*} 
\begin{center}
\begin{tabular}{cc}
\epsfig{file=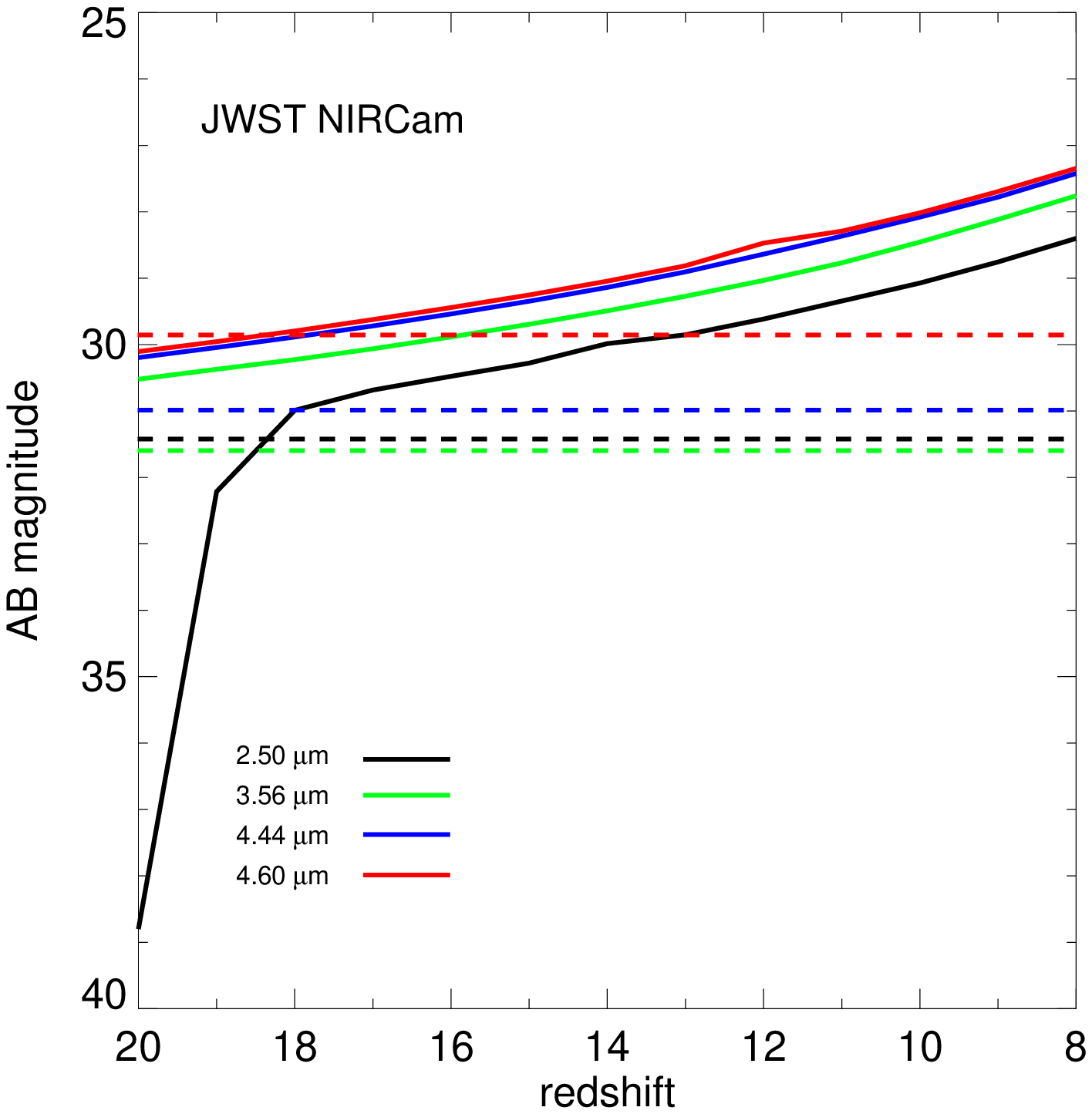,width=0.35\linewidth,clip=}  &
\epsfig{file=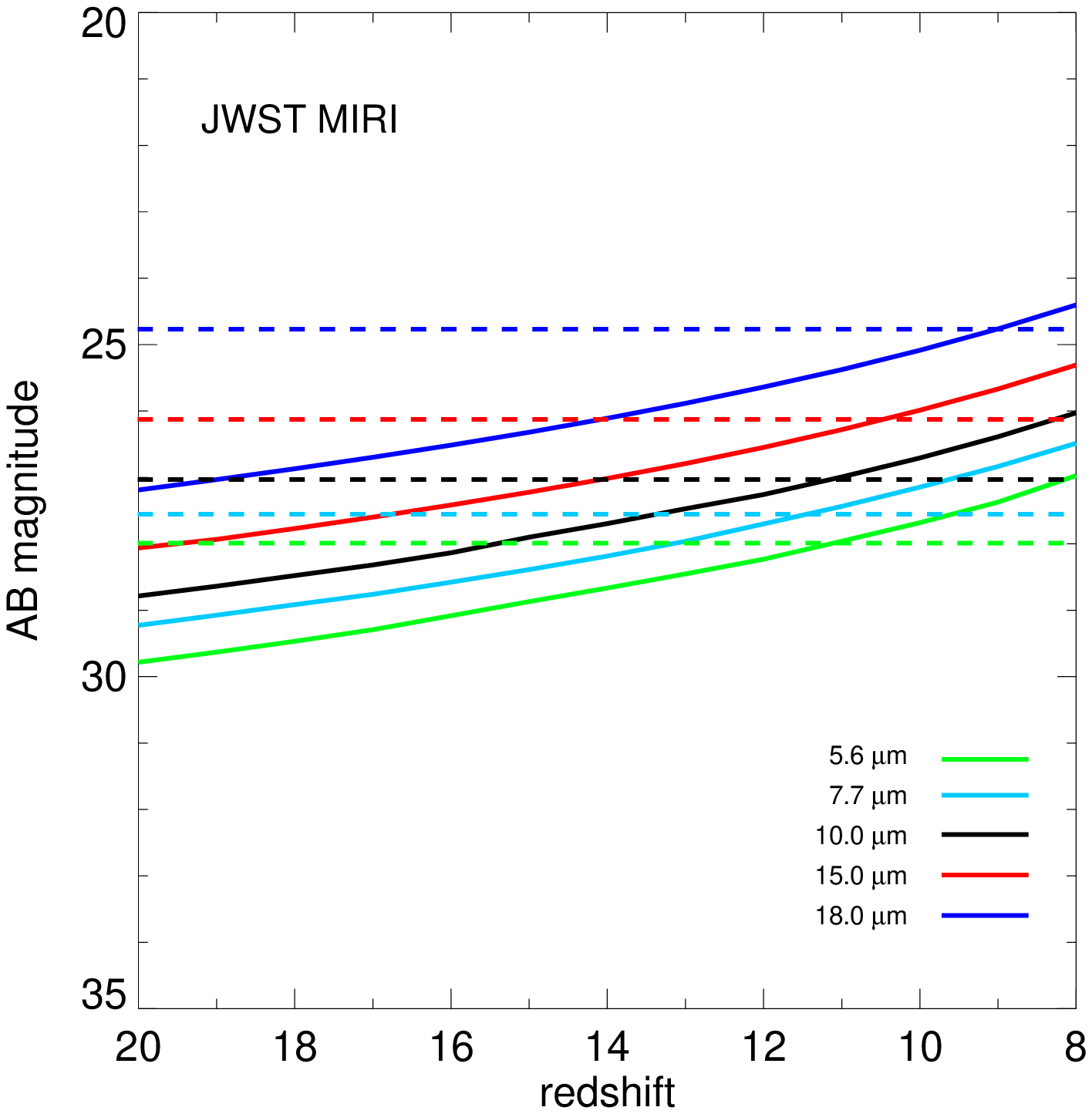,width=0.35\linewidth,clip=}  \\
\epsfig{file=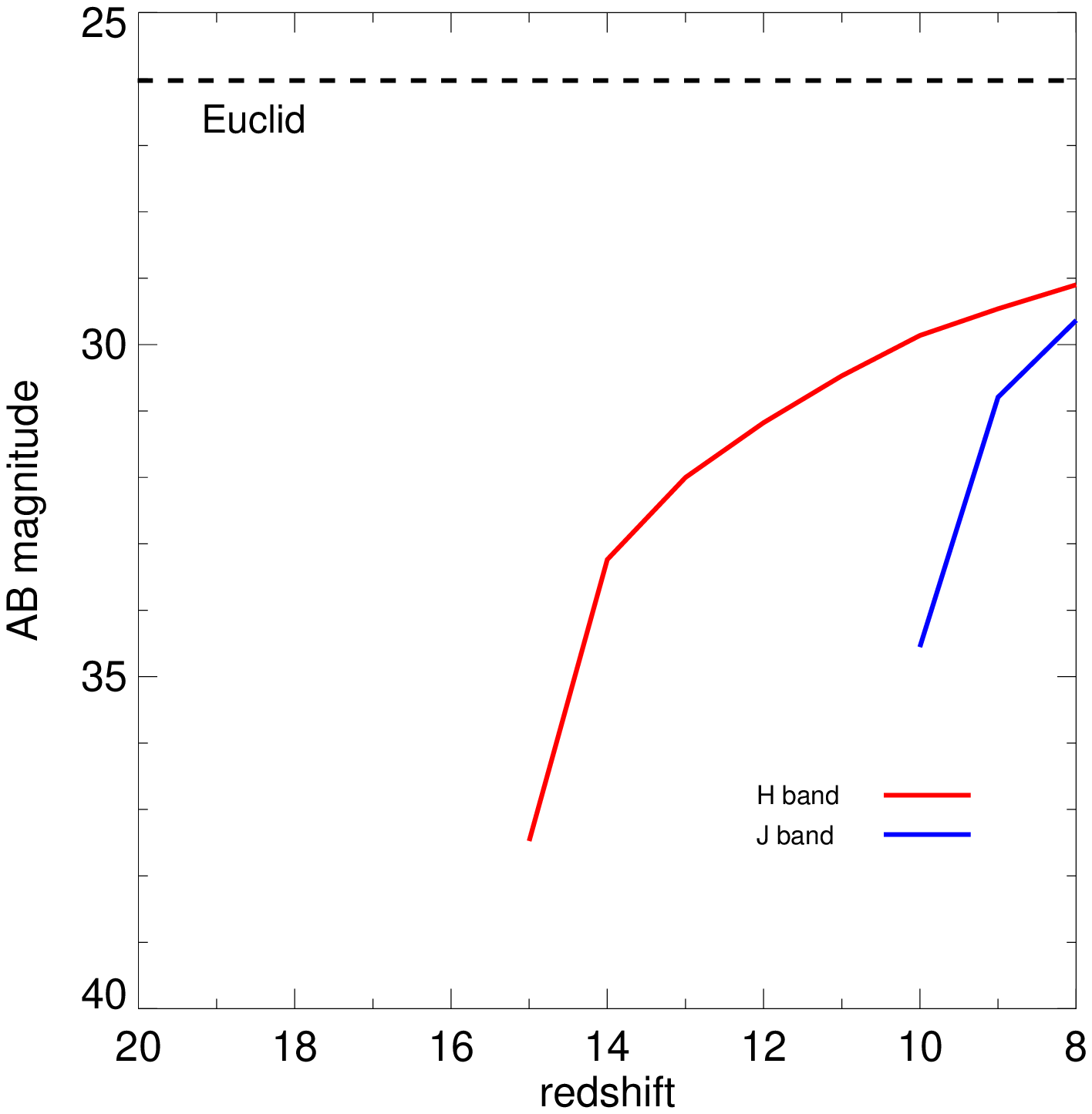,width=0.35\linewidth,clip=}  &
\epsfig{file=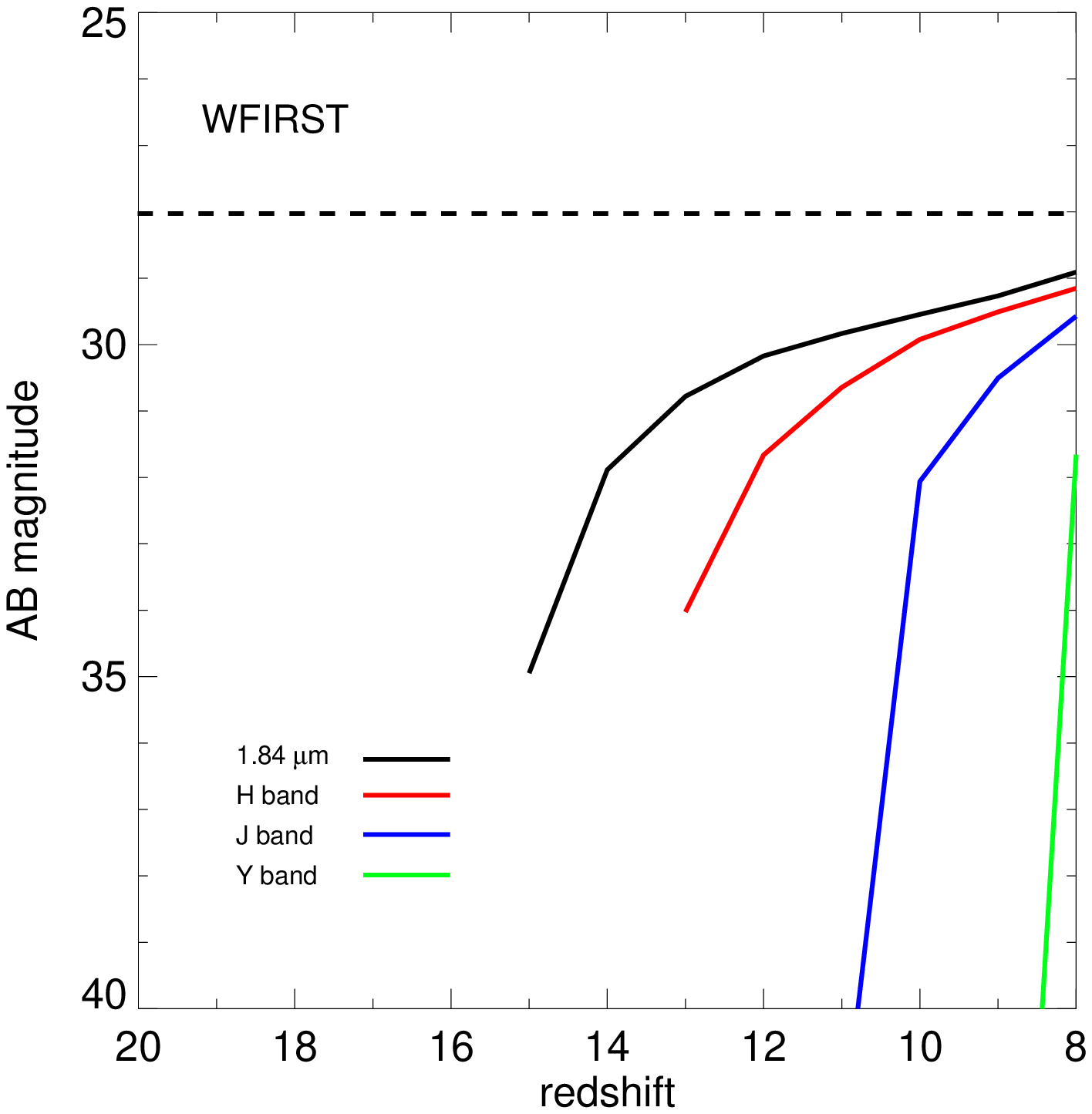,width=0.35\linewidth,clip=}  
\end{tabular}
\end{center}
\caption{NIR AB magnitudes for the 1.0 $\times$ 10$^5$ \Ms\ DCBH at birth as it would appear at $z =$ 8 - 20 in {\em JWST}, {\em Euclid} and WFIRST.  Top left: {\em JWST} NIRCam bands.  The horizontal dashed lines are 5$\sigma$ AB magnitude detection limits for 100 hour exposures in the filters of corresponding color (2.50 $\mu$m: 31.4 mag, 3.56 $\mu$m: 31.5 mag, 4.44 $\mu$m: 31.0 mag and 4.60 $\mu$m: 29.8 mag). Top right: {\em JWST} MIRI bands.  The horizontal dashed lines are 5$\sigma$ AB magnitude detection limits for 100 hour exposures in the filters of corresponding color (5.6 $\mu$m: 28.0 mag, 7.7 $\mu$m: 27.6 mag, 10.0 $\mu$m: 27.0 mag, 15.0 $\mu$m: 26.1 mag and 18.0 $\mu$m: 24.7 mag). Bottom left: {\em Euclid}.  Bottom right:  WFIRST.  The horizontal dashed lines in the bottom panels are detection limits for deep-drilling fields in Euclid and WFIRST (26 mag and 28 mag, respectively.)}
\vspace{0.1in}
\label{fig:ABmag} 
\end{figure*}

We show AB magnitudes for the DCBH at $z =$ 8 - 20 in {\em JWST} Near Infrared Camera (NIRCam) bands at 2.5 - 4.6 $\mu$m along with 5$\sigma$ detection limits for the filters for 100 hr exposures in the top left panel of Figure~\ref{fig:ABmag}.  The BH is clearly brighter in the NIR than its progenitor star \citep[see Figures 13, 4, and 3 of][respectively]{hos13,sur18a,sur19a}, with AB magnitudes that are 0.5 - 2.5 brighter depending on filter and wavelength.  The magnitudes in all four filters are also more tightly grouped together in consequence of the relatively flat power-law spectrum of the BH.  The drop in magnitude at $z =$ 18 at 2.5 $\mu$m is due to the redshifting of the Ly$\alpha$ absorption feature of the rest frame spectrum into that wavelength.  The BH is brightest in the 4.60 $\mu$m and 4.44 $\mu$m filters over all redshifts, with magnitudes that vary from 27.5 - 30.1 from $z = $ 8 - 20.  We find that detections in all four NIRCam filters are possible out to $z \sim$ 19 with 100 hr exposures and in all the bands redward of 3.56 $\mu$m out to $z \sim$ 25.

As shown in the upper right panel of Figure~\ref{fig:ABmag}, DCBH magnitudes in the {\em JWST} Mid-Infrared Instrument (MIRI) filters are significantly brighter than in NIRCam, ranging from 24.5 - 27 at 25.5 $\mu$m to 27 - 29.8 at 5.6 $\mu$m for $z =$ 8 - 20.  Some of these magnitudes are also much brighter than those of the progenitor star.  For example, the magnitudes of a red SMS vary from 31 - 32 at 5.6 $\mu$m over the same redshift range \citep{sur18a}.  The ordering of the magnitudes with filter wavelength for the SMS is opposite that of the DCBH, with the shortest wavelengths having the brightest magnitudes.  This feature is due to continuum absorption by H$^-$ in the envelope of the SMS that is absent from the host halo of the BH.  However, while the DCBH magnitudes are brighter in MIRI than NIRCam, the 5$\sigma$ detection limits for a 100 hour exposure are considerably dimmer, ranging from 24.8 at 18 $\mu$m to 28.0 at 5.6 $\mu$m.  They limit detections of DCBHs to $z =$ 9 at 18 $\mu$m to $z =$ 12 at 5.6 $\mu$m.  Nevertheless, these AB magnitudes reveal that MIRI could be a powerful instrument for the detection of DCBHs at high redshifts and could discriminate them from SMSs at the same epochs, for which there would be no MIRI signal.

We show DCBH magnitudes in the {\em Euclid} and WFIRST bands in the lower two panels of Figure~\ref{fig:ABmag}.  Absorption by the neutral IGM at $z \gtrsim$ 6 quenches flux in the Y, J and H bands at $z \gtrsim$ 7, 10 and 14, respectively, limiting DCBH detections to these redshifts in these filters.  Magnitudes vary from 29 - 34 in {\em Euclid} and 29 - 37 in WFIRST.  
The AB magnitude limits of 26 and 28 for surveys currently planned for {\em Euclid} and WFIRST, respectively, would rule out direct detections of DCBHs at $z \gtrsim$ 6 - 8.  We also computed AB magnitudes for all four cases with the spectrum for the 1$\times 10^5$ \Ms\ DCBH used in \citet{pac15} and found virtually no differences with those derived from our power-law spectrum.

\subsection{DCBH Formation / Detection Rates}

While our synthetic spectrum indicates that DCBHs would be detectable in multiband photometric surveys with {\em JWST} at $z \sim$ 8 - 20, the prospect of actually finding such objects in a given survey depends on their formation rates and the time interval over which a DCBH is likely to display tell-tale spectral or photometric signatures.  \citet{wise19} and \citet{regan19} identified atomically cooling halos at $z \gtrsim 12$ in the Renaissance simulations that could form DCBHs.  The 3 DCBH candidate halos that appeared in their 220 cMpc$^3$ average-density region over the $\sim 70$ Myr from $z \sim$ 14 - 12 imply a formation rate of $\sim 10^{-10}$ cMpc$^{-3}$ yr$^-1$ at these redshifts.  While their simulations did not track the subsequent evolution of the gas in these halos, this formation rate can be used to place an upper limit on detections of DCBHs in future {\em JWST} surveys. 

If we adopt a characteristic time scale of $10^7$ yr for the validity of our spectrum (set by when star formation likely begins in its host halo) then we expect a comoving density of observable DCBHs of $\sim 10^{-3}\ f_\mathrm{DCBH}$ cMpc$^{-3}$, where $f_\mathrm{DCBH}$ is the fraction of candidate halos that produce $\sim 10^5\ M_\odot$ BHs. The {\em JWST} NIRcam field of view (9.7 arcmin$^2$) covers $1.3\times 10^4$ cMpc$^3$ per unit redshift at $z \sim 12$, so one would expect $\sim 10 f_\mathrm{DCBH}$ detectable DCBHs for each such survey field.  With planned medium-deep NIRCam multiband surveys covering $\sim 20$ times this area down to AB mag 29 in the longest-wavelength NIRCam filters \citep{rieke19}, the prospects for detecting DCBHs photometrically with {\em JWST} would appear to be quite good, even if just some minor fraction ($f_\mathrm{DCBH}\gtrsim 0.01$) of the \citet{regan19} candidate halos end up forming them.  Another route to detection could be to search the field around an unusually bright $z \sim$ 15 galaxy found by some other means, as \citet{wise19} and \citet{regan19} note that the formation rate of DCBHs may rise by more than an order of magnitude in high-density regions, where the most massive first galaxies are also expected to form.

\section{Discussion and Conclusion}

With NIRCam AB mag photometry limits of 31 - 32 and NIRSpec limits of $\sim$ 29, {\em JWST} will be able to detect the birth of the first quasars at $z \gtrsim$ 20 and spectroscopically confirm their redshift out to $z \sim$ 10 - 12.  Our DCBH magnitudes are consistent with simplified 1D calculations in past studies \citep{yue13,pac15,nat17}.  As shown in the previous section, up to 10 DCBHs could appear in any given {\em JWST} survey field from $z =$ 8 - 20.  But the prospects for discovering them would be better if they could also be found by {\em Euclid} and WFIRST because their wide fields would enclose far more of them at high redshifts.  Once flagged, DCBH candidates could then be examined with {\em JWST} or ground-based extremely large telescopes in greater detail.  But, as shown in Figure \ref{fig:ABmag}, DCBH magnitudes in the H band magnitudes at $z =$ 8 - 20 are dimmer than the detection limits currently envisioned for {\em Euclid} and WFIRST (26 and 28, respectively).  In principle, these magnitudes could become brighter if accretion rates exceed the Eddington limit but only modestly so because the luminosity rises only logarithmically with such rates, not linearly.

However, this does not mean {\em Euclid} and WFIRST cannot find DCBHs at birth because only modest gravitational lensing is required to boost their fluxes above their detection limits.  The survey areas of both missions will enclose thousands of galaxy clusters and massive galaxies that could lense flux from background DCBHs, and at $z \sim$ 8 - 14 magnification factors of only 10 - 100 would be required to reveal them.  It is likely that a sufficient fraction of their survey areas will be magnified by such factors \citep{ryd20a,fa19}.  Even higher magnifications may be realized in future surveys of individual cluster lenses by {\em JWST} but at the cost of smaller lensing volumes \citep[e.g.,][]{wet13c,wind18}. 

As discussed earlier, the host halo of our DCBH is constantly replenished by new gas from cold accretion flows, in contrast to most other atomically-cooled halos at high redshift.  The lower average densities of those halos would have less effect on the emergent spectrum of the BH and they would therefore have somewhat lower fluxes in the NIR.  DCBHs in general can be distinguished from their SMS progenitors at high redshift because they are brighter and have much higher ratios of flux in the MIRI and NIRCam bands.  \citet{nat17} found that DCBHs can be readily distinguished from early galaxies at similar redshifts in color-color space, which would also be true for our models because of the similarities in source spectra of both studies \citep[see also][]{val18}.  Also, unlike SMSs and high-$z$ galaxies, they are transients because of variations in cosmological flows onto them on timescales as short as the redshifted light-crossing time of the BH.  Periodic dimming and brightening could therefore tag these objects as high-$z$ BHs in transient surveys with well-chosen cadences proposed for {\em JWST} such as FLARE \citep{flare}.  Synergies between {\em Euclid} or WFIRST and {\em JWST} or 20$+$ m ground-based telescopes could open the era of $z =$ 8 - 20 quasar astronomy in the coming decade.

\acknowledgments

The authors thank the anonymous referee for constructive comments that improved the quality of this paper.  D. J. W. was supported by the Ida Pfeiffer Professorship at the Institute of Astrophysics at the University of Vienna and by STFC New Applicant Grant ST/P000509/1.  E. Z. acknowledges funding from the Swedish National Space Board.  F. P. acknowledges support from a Clay Fellowship administered by the Smithsonian Astrophysical Observatory and from the Black Hole Initiative at Harvard University, which is funded by grants from the John Templeton Foundation and the Gordon and Betty Moore Foundation.  M. H. acknowledges financial support from the Carlsberg Foundation via a Semper Ardens grant (CF15-0384).  Our simulations were performed on the Sciama cluster at the Institute of Cosmology and Gravitation at the University of Portsmouth.  We also acknowledge support by the state of Baden-W{\"u}rttemberg through bwHPC (the bwForCluster).

\bibliographystyle{apj}
\bibliography{refs}

\end{document}